Evolution of Superconductivity in BiS$_2$-Based Superconductor LaO$_{0.5}$F$_{0.5}$Bi(S$_{1-x}$Se$_x$)$_2$

Takafumi Hiroi, Joe Kajitani, Atsushi Omachi, Osuke Miura, and Yoshikazu Mizuguchi

Department of Electrical and Electronic Engineering, Tokyo Metropolitan University, Hachioji, Tokyo 192-0397, Japan

Abstract

We have systematically investigated the crystal structure, magnetic susceptibility, and electrical resistivity of the BiS$_2$-based superconductor LaO$_{0.5}$F$_{0.5}$Bi(S$_{1-x}$Se$_x$)$_2$ ($x = 0 - 0.7$). With expanding lattice volume by Se substitution, bulk superconductivity was induced for $x \geq 0.2$, and the highest $T_c$ of 3.8 K was observed in $x = 0.5$ (LaO$_{0.5}$F$_{0.5}$BiSSe). Metallic conductivity was observed for $x \geq 0.3$ in the resistivity measurement, whereas semiconducting-like behavior was observed for $x \leq 0.2$. The induction of bulk superconductivity by the partial substitution of S by Se in the LaO$_{0.5}$F$_{0.5}$BiS$_2$ superconductor should be positively linked to the enhancement of metallic conductivity.

1. Introduction

Since the discovery of the novel layered $BiS_2$-based superconductor $Bi_4O_4S_3$ [1], many studies exploring new superconductors and clarifying the mechanisms of superconductivity in the $BiS_2$ family have been carried out. Thus far, three types of $BiS_2$-based superconductor, $Bi_4O_4S_3$, $RE$OBiS$_2$ ($RE$ = La, Ce, Pr, Nd, Sm, and Yb) [2-9], and $AE$FBiS$_2$ ($AE$ = Sr, and Eu) [10-14] have been discovered, and the highest transition temperature ($T_c$) reaches about 11 K in $LaO_{0.5}F_{0.5}BiS_2$ [15]. $BiS_2$-based superconductors have a layered structure consisting of the alternate stacking of the $BiS_2$ superconducting layer and the blocking layer. The superconducting properties of the $BiS_2$ family can be tuned by substituting the elements at the blocking layers. These characteristics resemble those of cuprate and Fe-based superconductors [16,17]. In $BiS_2$-based layered materials, electron doping into the $BiS_2$ layers is essential for the induction of superconductivity [18]. For example, in the $LaOBiS_2$ system, superconductivity is induced by a partial substitution of $O^{2-}$ by $F^-$ or by a partial substitution of $La^{3+}$ by $M^{4+}$ at the blocking layers [2,19]. Namely, element substitution at the blocking layer is effective for inducing superconductivity and tuning its superconducting properties in $BiS_2$-based superconductors. Furthermore, several high-pressure studies revealed that superconducting properties are sensitive to changes in local crystal structures: the $T_c$ can be enhanced by optimal lattice contraction [20-27]. In addition, it was reported that the chemical pressure effect could enhance superconducting properties in the $RE$(O,F)BiS$_2$ system [8,28-31]. In $Ce_{1-x}Nd_x$(O,F)BiS$_2$ and $Nd_{1-x}Sm_x$(O,F)BiS$_2$, bulk superconductivity was induced by the decrease in lattice volume, and the highest $T_c$ of 5.6 K was observed in $Nd_{0.2}Sm_{0.8}O_{0.7}F_{0.3}BiS_2$ [29]. In $La_{1-x}Sm_xO_{0.5}F_{0.5}BiS_2$, bulk superconductivity was induced by the chemical pressure effect, and the highest $T_c$ of 5.4 K was obtained for La$_{0.2}$Sm$_{0.8}$O$_{0.5}$F$_{0.5}$BiS$_2$ [31]. These experimental results suggest that the intrinsic (bulk and higher $T_c$) superconductivity could be induced by optimizing crystal structures in the BiS$_2$-based family. In fact, the importance of the optimization of the local structure in determining superconducting or physical properties was suggested in theoretical studies [32,33] and X-ray absorption spectroscopy studies [34,35].

There are many reports on the element substitution effects of blocking layers in REO$_{1-x}$F$_x$BiS$_2$. On the other hand, there are only a few reports on the effect of element substitution within the superconducting layer. In Bi$_4$O$_4$S$_3$, a partial substitution of S by Se decreased $T_c$ [36]. The $T_c$ of Nd(O,F)BiS$_2$ decreased with a partial substitution of S by Se as well [37]. In contrast, the observation of bulk superconductivity with a $T_c$ of 3.8 K in LaO$_{1-x}$F$_x$BiSSe, in which 50% of the S sites are replaced by Se, has recently been reported [38]. In this study, the effect of Se substitution at the superconducting layers on the superconducting properties of LaO$_{0.5}$F$_{0.5}$Bi(S$_{1-x}$Se$_x$)$_2$ were systematically investigated as a function of Se concentration. We synthesized polycrystalline samples of Se-substituted LaO$_{0.5}$F$_{0.5}$Bi(S$_{1-x}$Se$_x$)$_2$, investigated their physical properties, and established a superconductivity phase diagram.

2. Experimental Procedure

The polycrystalline samples of LaO$_{0.5}$F$_{0.5}$Bi(S$_{1-x}$Se$_x$)$_2$ ($x$ = 0 - 0.7) were prepared by a solid-state reaction method. First, powders of Bi$_2$Se$_3$ were presynthesized from Bi (99.999%) and Se (99.99%) grains. The weighed Bi and Se grains were sealed in an evacuated quartz tube and heated at 800 °C for 10 h. The starting materials for LaO$_{0.5}$F$_{0.5}$Bi(S$_{1-x}$Se$_x$)$_2$ were La$_2$O$_3$ powder (99.9%), LaF$_3$ powder (99.9%), La$_2$S$_3$ powder (99.9%), $Bi_2Se_3$ powder, Bi grains, and S grains (99.9%). The mixtures with nominal compositions of $LaO_{0.5}F_{0.5}Bi(S_{1-x}Se_x)_2$ were mixed well, ground, pelletized, sealed in an evacuated quartz tube and annealed at 700 °C for 15 h. The calcined samples were ground, pelletized, and heated in an evacuated quartz tube by the same process as the first heating. The obtained samples were characterized by powder X-ray diffraction (XRD) analysis using the $\theta$-$2\theta$ method with CuKα radiation. The lattice constants were estimated by the Rietveld method using RIETAN-FP [39]. The temperature dependence of magnetic susceptibility was measured using a superconducting quantum interface device (SQUID) magnetometer with an applied field of 5 Oe after both zero-field cooling (ZFC) and field cooling (FC). The temperature dependence of resistivity was measured by the four-probe method.

3. Results and Discussion

Figure 1 shows the powder XRD patterns of $LaO_{0.5}F_{0.5}Bi(S_{1-x}Se_x)_2$ ($x$ = 0 - 0.7). For all the samples, the obtained XRD peaks of $LaO_{0.5}F_{0.5}BiS_2$ could be indexed using the *P4/nmm* space group as indicated by Miller indices displayed with the XRD pattern of $x$ = 0. The Rietveld refinement result for the XRD pattern of $LaO_{0.5}F_{0.5}BiSSe$ ($x$ = 0.5) is shown in Fig. 2. Except for the tiny peak of bismuth oxide at approximately $2\theta$ = 27.5º, almost all the observed profiles were refined, indicating that the S/Se solution system is single phase ($R_{wp}$ ~ 0.16). Since $LaO_{0.5}F_{0.5}Bi(S_{1-x}Se_x)_2$ is a solid-solution system, we carried out refinements by inputting a virtual element of S/Se: the mixing ratio was fixed to the starting nominal compositions. Then, the Rietveld refinement was carried out using typical crystal structure data of the $REOBiS_2$-type materials. Having considered the resolution of our laboratory-level XRD analysis with powder samples containing the S/Se solution, we simply discuss the variations in the lattice constants of $a$-axis, $c$-axis, and volume for the $LaO_{0.5}F_{0.5}Bi(S_{1-x}Se_x)_2$ system in the crystal structure part of this article.

Figure 3 shows the Se concentration dependences of the lattice constants throughout (a) $a$-axis, (b) $c$-axis and (c) volume. The lattice constants of $a$-axis and $c$-axis increase with increasing Se concentration. These changes are consistent with the difference in ionic radius between $S^{2-}$ (184 pm) and $Se^{2-}$ (198 pm). The $a$-axis monotonically increases with increasing Se concentration. The $c$-axis slightly increases up to $x = 0.4$ and largely increases above $x = 0.5$. These tendencies indicate that the Se ions are selectively introduced into one site of the $BiS_2$ superconducting layer, although we cannot obtain further information within the present analysis.

Figure 4(a) shows the temperature dependences of the ZFC magnetic susceptibility ($4\pi\chi$) with an applied magnetic field of 5 Oe for $LaO_{0.5}F_{0.5}Bi(S_{1-x}Se_x)_2$. For $x = 0$ and 0.1, small diamagnetic signals were observed, indicating that the shielding volume fractions were clearly as low as superconducting states of a bulk superconductor. We regard these two samples as *filamentary* superconductors. $T_c$ was defined as the irreversible temperature ($T_c^{irr}$) that is the bifurcation point between the ZFC and FC susceptibility curves, as shown in Fig. 4(b). $T_c$ for $x = 0$ is 2.5 K. $T_c$ decreases to 2.2 K at $x = 0.1$. Above $x = 0.2$, $T_c$ increases, and the shielding volume fraction is clearly enhanced. On the basis of the large shielding fraction, we regard the samples with $x = 0.2 - 0.7$ as *bulk* superconductors. In fact, a partial substitution of S by Se induces bulk superconductivity in the $LaO_{0.5}F_{0.5}Bi(S_{1-x}Se_x)_2$ system. The highest $T_c$ of 3.8 K is observed for $x = 0.5$. For $x > 0.5$, $T_c$ decreases with increasing Se concentration.

Figure 5(a) shows the temperature dependences of electrical resistivity for LaO$_{0.5}$F$_{0.5}$Bi(S$_{1-x}$Se$_x$)$_2$ in the temperature range between 1.8 and 300 K. Figure 5(b) is the enlargement magnified view of Fig. 5(a) at low temperatures near $T_c$. For LaO$_{0.5}$F$_{0.5}$BiS$_2$, semiconducting-like (bad-metal-like) behavior is observed above its $T_c$, and zero resistivity is observed at 2.5 K. Resistivity clearly decreases with increasing Se concentration, and the semiconducting-like behavior disappears for $x \geq 0.3$. For $x = 0.3$, resistivity is almost constant as a function of temperature below 300 K. For $x \geq 0.4$, the temperature dependence of resistivity shows metallic behavior. In fact, the values of resistivity at room temperature for $x = 0.4 - 0.7$ are 5 times as low as that for $x = 0$. The estimated zero-resistivity temperatures ($T_c^{zero}$) for all the samples correspond to the $T_c^{irr}$ estimated by the magnetic susceptibility measurements. These results indicate that the evolution of bulk superconductivity should be related to the enhancement of metallic conductivity in the Se-substituted LaO$_{0.5}$F$_{0.5}$Bi(S$_{1-x}$Se$_x$)$_2$ system.

Figure 6 shows the superconductivity phase diagram of the LaO$_{0.5}$F$_{0.5}$Bi(S$_{1-x}$Se$_x$)$_2$ system where $T_c$ ($T_c^{irr}$) is plotted as a function of Se concentration ($x$). With increasing Se concentration, $T_c$ firstly decreases to 2.2 K at $x = 0.1$. $T_c$ increases with increasing Se concentration within $0.2 \leq x \leq 0.5$ and reaches the highest $T_c$ of 3.8 K at $x = 0.5$. In the Se-poor region, the samples show *filamentary* superconductivity, and bad-metal-like conductivity is observed. With increasing Se concentration, metallic conductivity is induced, and superconducting properties are enhanced. The appearance of bulk superconductivity should be linked to the enhancement of metallic characteristics.

Although we have revealed the superconductivity phase diagram of the LaO$_{0.5}$F$_{0.5}$Bi(S$_{1-x}$Se$_x$)$_2$ system as a function of Se concentration, we should mention that the highest $T_c$ in this system is clearly lower than 11 K for the high-pressure phase of LaO$_{0.5}$F$_{0.5}$BiS$_2$. For the high-pressure phase of LaO$_{0.5}$F$_{0.5}$BiS$_2$, Tomita et al. suggested that the crystal structure is not tetragonal but monoclinic [23]. Namely, the higher-$T_c$ phase would appear in a distorted structure. In this study, we have revealed that the tetragonal structure does not change with Se substitution up to $x = 0.7$. On the basis of these facts, we assume that the partial substitution of S by Se could enhance superconductivity, but it could not induce the higher-$T_c$ phase as those revealed in previous high-pressure studies of BiS$_2$-based superconductors [21,22,26,27].

4. Summary

We have systematically investigated the crystal structure, magnetic susceptibility, and electrical resistivity of polycrystalline samples of LaO$_{0.5}$F$_{0.5}$Bi(S$_{1-x}$Se$_x$)$_2$. With increasing Se concentration, the lattice constants $a$-axis and $c$-axis systematically increased. With expanding lattice volume by Se substitution, bulk superconductivity was induced for $x \geq 0.3$, and the highest $T_c$ of 3.8 K was observed for $x = 0.5$. One of the notable facts is that the metallic conductivity was observed for $x \geq 0.3$ in the resistivity measurement, whereas bad-metal-like (semiconducting-like) behavior was observed for $x \leq 0.2$. The induction of bulk superconductivity by the partial substitution of S by Se in the LaO$_{0.5}$F$_{0.5}$BiS$_2$ superconductor should be positively linked to the enhancement of metallic conductivity.

Acknowledgements
We thank Dr. Y. Takano, Mr. T. Yamaki, and Dr. M. Tanaka of the National Institute for Materials Science and N. L. Saini of Sapienza University of Rome for fruitful discussion. This work was partly supported by JSPS KAKENHI Grant Numbers 25707031 and 26600077.

Figure captions

Fig. 1. (Color online) Powder XRD patterns of LaO$_{0.5}$F$_{0.5}$Bi(S$_{1-x}$Se$_x$)$_2$ ($x$ = 0 - 0.7).

Fig. 2. (Color online) Powder XRD pattern and Rietveld refinement result for LaO$_{0.5}$F$_{0.5}$BiSSe ($x$ = 0.5).

Fig. 3. (Color online) Se concentration dependences of the lattice constants (a) $a$-axis, (b) $c$-axis, and (c) volume for LaO$_{0.5}$F$_{0.5}$Bi(S$_{1-x}$Se$_x$)$_2$ ($x$ = 0 - 0.7).

Fig. 4. (Color online) Temperature dependences of the ZFC magnetic susceptibilities (4$\pi\chi$) of LaO$_{0.5}$F$_{0.5}$Bi(S$_{1-x}$Se$_x$)$_2$ ($x$ = 0 - 0.7). (b) Temperature dependence of magnetic susceptibility for $x$ = 0.5 at which the onset of the superconducting transition is enlarged.

Fig. 5. (Color online) (a) Temperature dependences of the electrical resistivity of LaO$_{0.5}$F$_{0.5}$Bi(S$_{1-x}$Se$_x$)$_2$ ($x$ = 0 - 0.7). (b) Enlargement of the temperature dependences of the electrical resistivity of LaO$_{0.5}$F$_{0.5}$Bi(S$_{1-x}$Se$_x$)$_2$ ($x$ = 0.2 - 0.7).

Fig. 6. (Color online) Superconductivity phase diagram of LaO$_{0.5}$F$_{0.5}$Bi(S$_{1-x}$Se$_x$)$_2$ ($x$ = 0 - 0.7). $T_c$ is plotted as a function of Se concentration. On the basis of the values of -4$\pi\chi$ (ZFC) at 2 K, it is clear that bulk superconductivity is induced within the range of 0.2 $\leq$ $x$ $\leq$ 0.7. Metallic conductivity is observed for $x$ $\geq$ 0.3, while bad-metal-like (semiconducting-like) behavior is observed for $x \leq$ 0.2.

Fig. 1

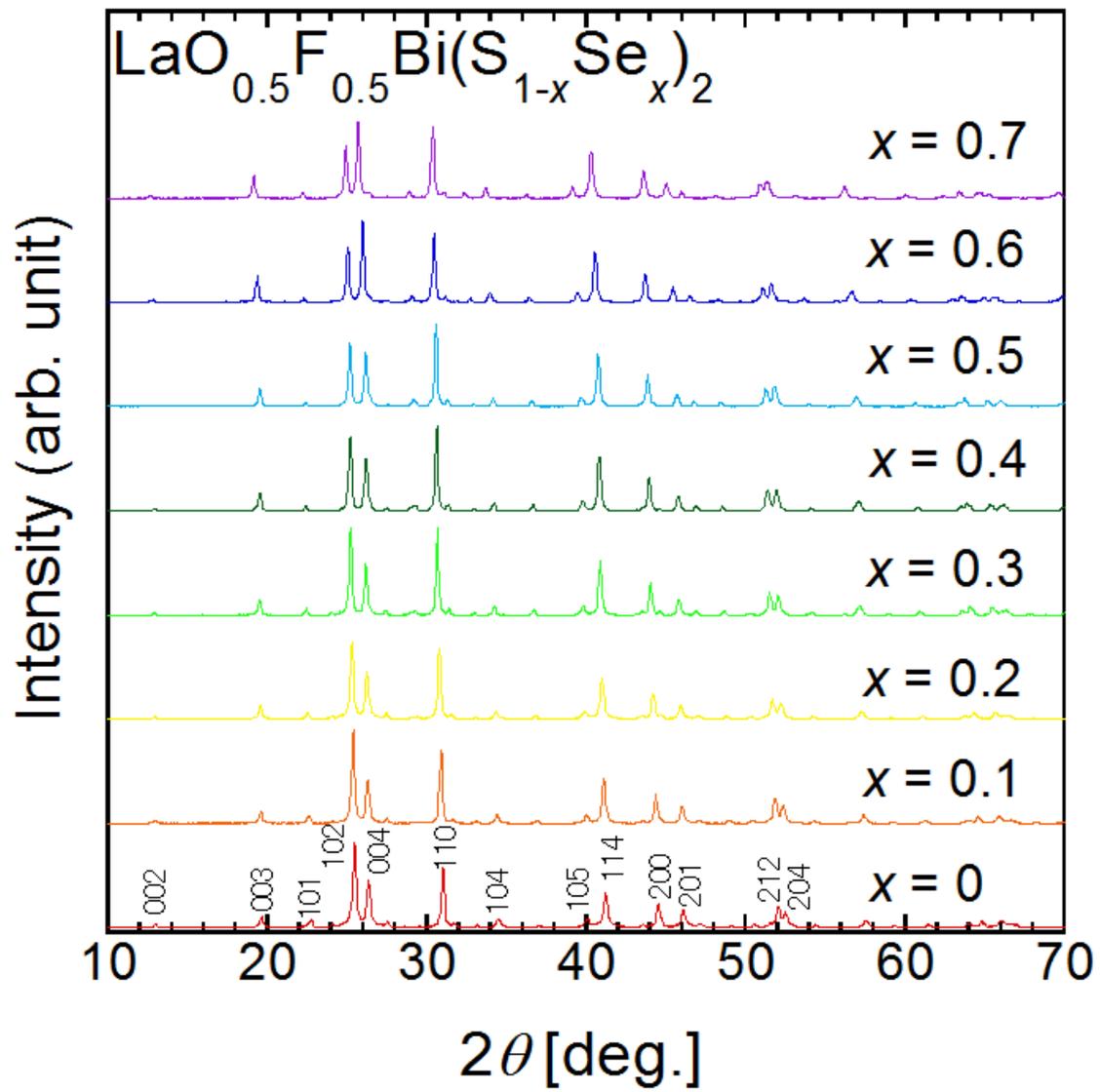

Fig. 2

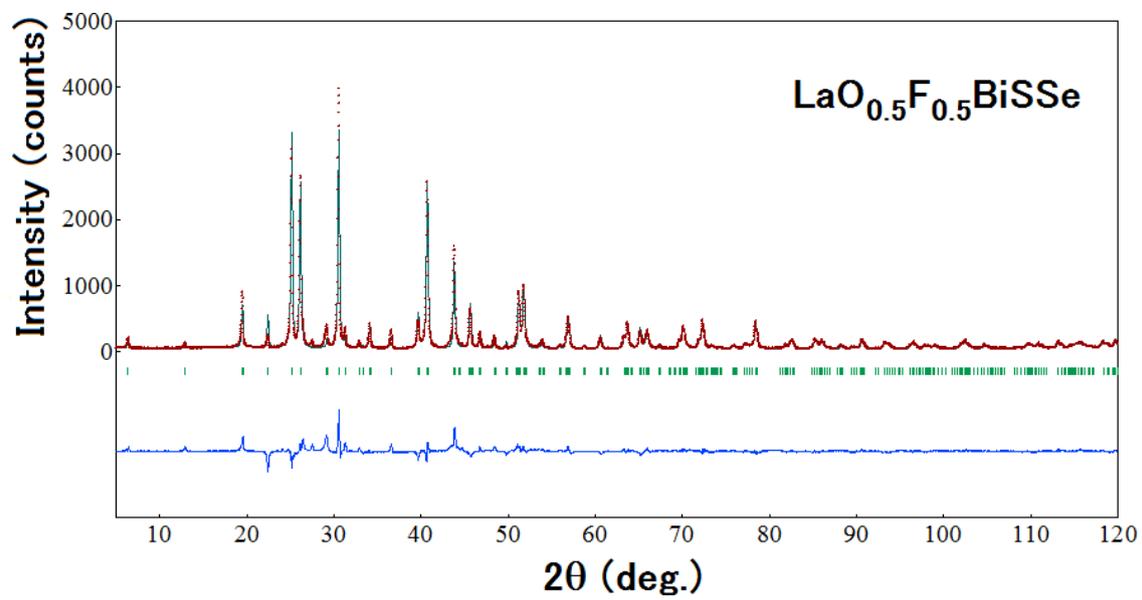

Fig. 3

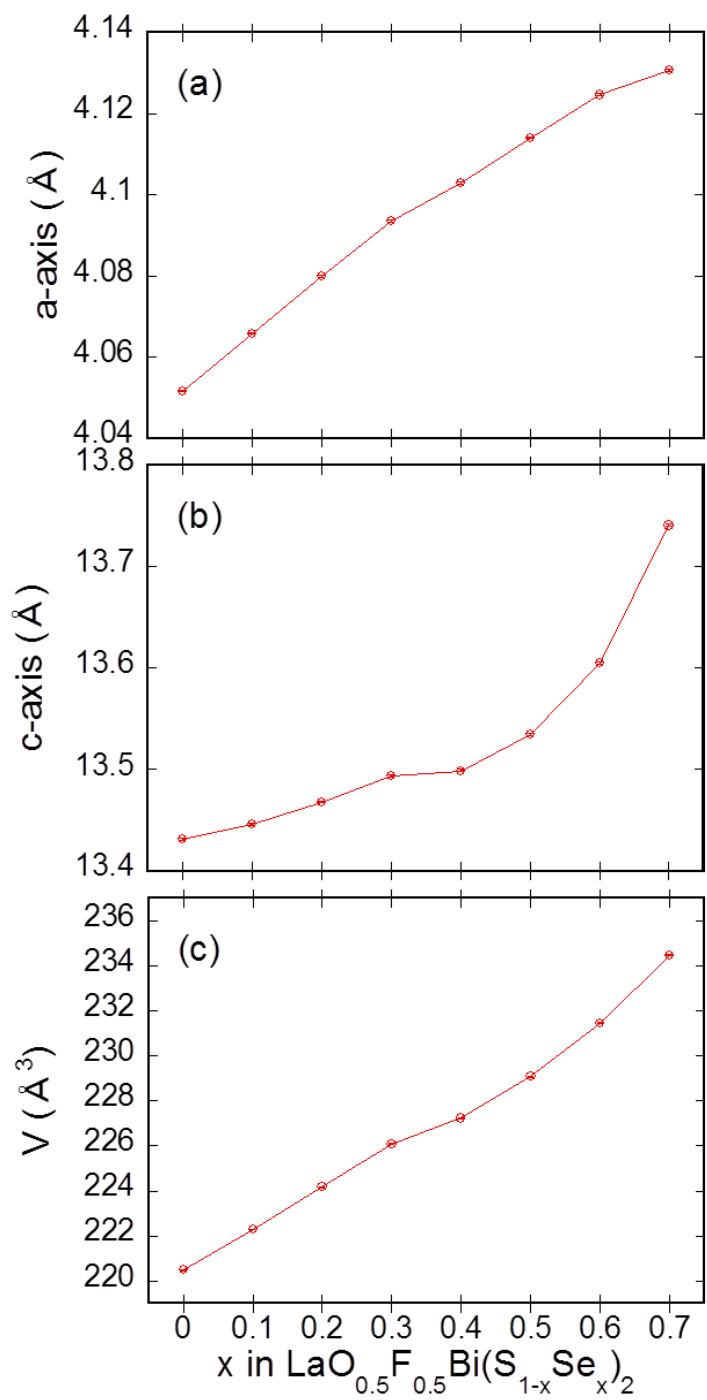

Fig. 4

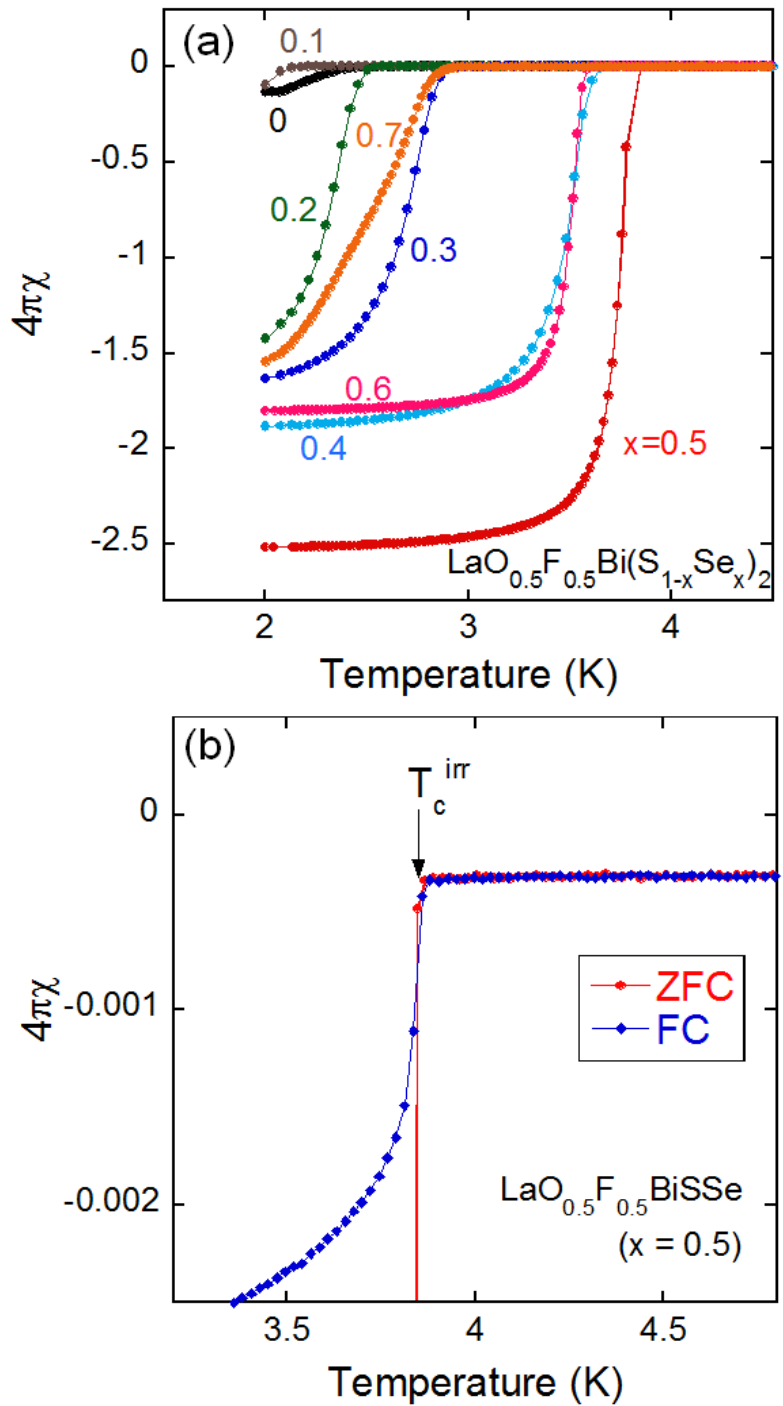

Fig. 5

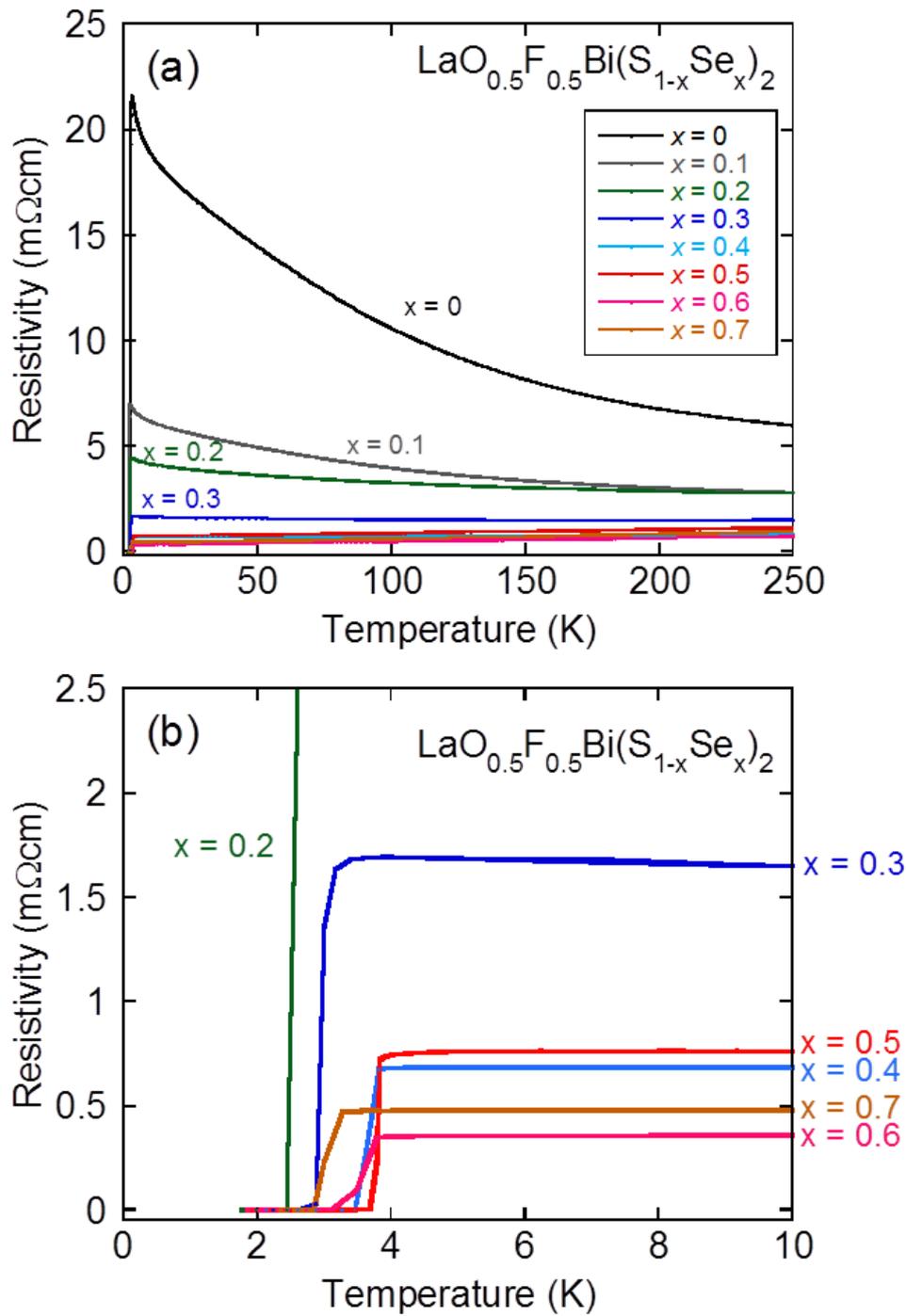

Fig. 6

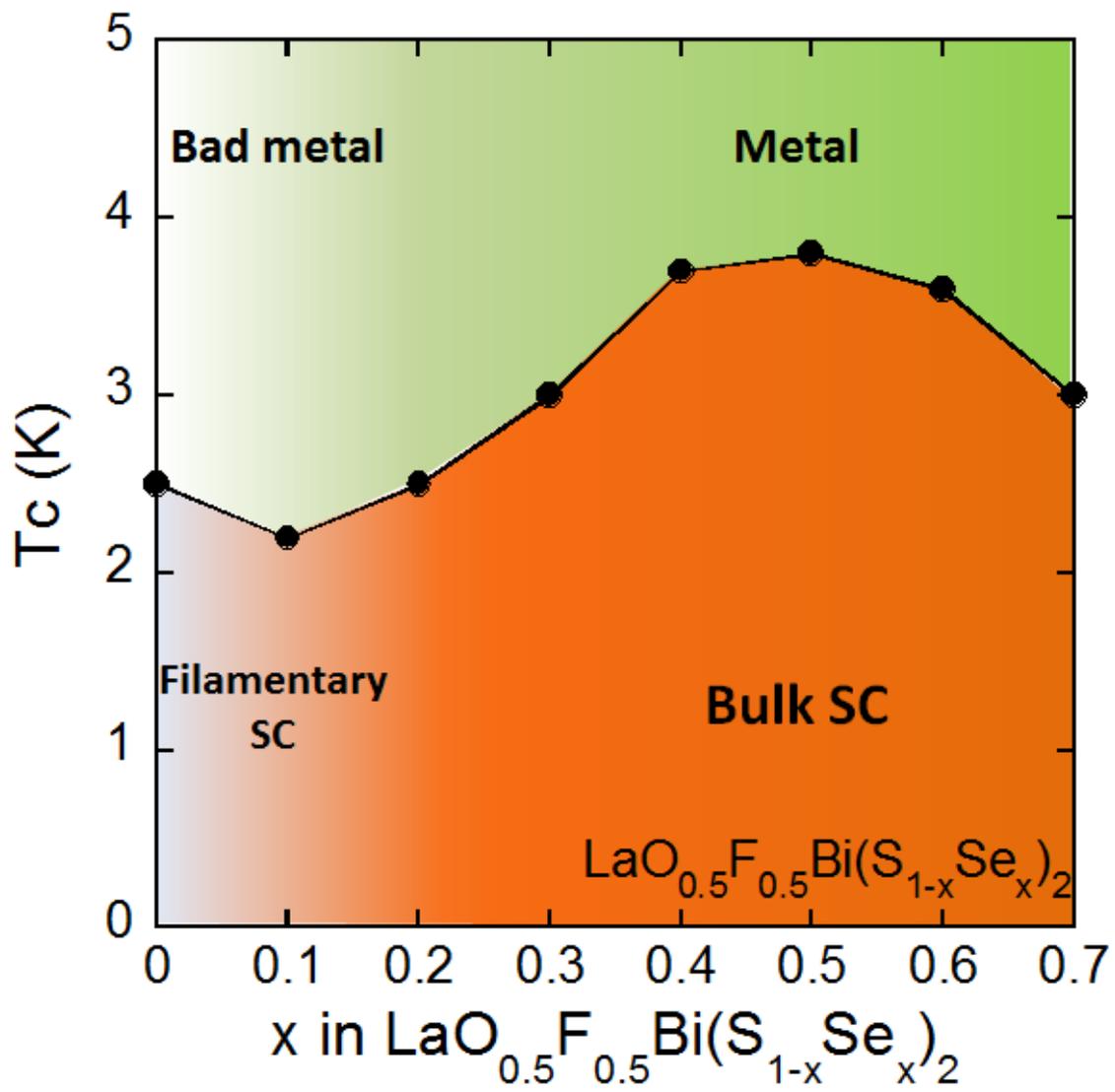